\def\ref#1{$[#1]$}
\vsize 21.6 truecm
\hsize 15.2 truecm

\font\twelvebf=cmbx12
\font\eightbf=cmbx8
\font\eightrm=cmr8

\def\newline{\hfill\break}

\noindent{\twelvebf  STATISTICAL MECHANICS AND ERROR-CORRECTING CODES }

\vglue 1.5 true cm
\centerline{{\bf  Nicolas Sourlas}} 
\medskip
\centerline{Laboratoire de Physique Th\'eorique de l' Ecole Normale 
Sup\'erieure  \footnote {*}  
{Unit\'e Propre du Centre National de la Recherche Scientifique, 
associ\'ee \`a l' Ecole Normale Sup\'erieure et \`a l'Universit\'e de 
Paris-Sud. } }
\centerline{ 24 rue Lhomond, 75231 Paris CEDEX 05, France. }
\centerline { e-mail: $\ $ sourlas@physique.ens.fr }

{\vskip 24pt}

\noindent{\eightbf Abstract:} {\eightrm I will show that there is a deep relation between error-correction 
codes and certain mathematical models of spin glasses. In particular minimum error 
probability decoding is equivalent to finding the ground state 
 of the corresponding spin system.  
The most probable value of a symbol is related to the magnetization 
at a different temperature. Convolutional codes correspond to one-dimensional 
spin systems and Viterbi's decoding algorithm to the transfer matrix 
algorithm of Statistical Mechanics. A particular spin-glass model, 
which is exactly soluble, 
corresponds to an ideal code, i.e. a code which allows error-free 
communication if the rate is below channel capacity.
}

%{\vskip 3pt}

%\noindent{\eightbf Keywords:} {\eightrm List of keywords.}

{\vskip 12pt}

%\noindent{\twelvebf 1 \quad Introduction or

%\centerline{\bf }
%\medskip 
%\centerline{{\bf STATISTICAL MECHANICS AND ERROR-CORRECTING CODES }}
%\vglue 1.5 true cm
%\centerline {{\bf Abstract }}
%\vglue .8 true cm

%\vglue 1.5 true cm
%\noindent LPTENS 98/ 43
%\medskip
%\noindent November 98

%\vfill\eject

 The mathematical theory of communication\ref{1,2} is probabilistic in 
nature. Both the production of information and its transmission 
 are considered as probabilistic events. A source is producing 
information messages according to a certain 
probability distribution. 
 Each message consists of a sequence of $N$ bits 
$\vec \sigma = \{ \sigma_{1}, \cdots , \sigma_{N} \}, \ \sigma_{i} = \pm 1 $
 and it is assumed 
 that the probability $P_{s} (\vec \sigma )  \equiv  
\exp -H_{s}(\vec \sigma ) $ of any particular sequence $\vec \sigma $ 
is known. According to Shannon the information content of the message is 
$ -\ln P_{s} (\vec \sigma ) $ and the average information of the source 
is given by 
$$ - \sum_{\vec \sigma } P_{s} (\vec \sigma ) \ln P_{s} (\vec \sigma ) $$
 The messages are sent through 
a transmission channel. In general there is noise 
during transmission (which may have different origins) 
which corrupts the transmitted message.
 If a $ \sigma  = \pm 1 $ is sent through the transmission 
channel, because of the noise, the output will be a real number 
$ u$, in general different from $ \sigma $. Again, the statistical properties 
of the transmission channel are supposed to be known. Let us call 
$ Q ( \vec u | \vec \sigma ) du $ the probability for the transmission 
channel's output to be between $ u $ and $ u + du$,
 when the input was $ \sigma $. 
$ Q ( \vec u | \vec \sigma ) $  is supposed to be known. 
Because of the noise during the transmission, there is a loss of 
information. The channel capacity $ {\cal C } $ 
is defined as the maximum information per unit time which 
 can be transmitted through the channel. The maximum is taken 
over all possible sources.

Thanks to Shannon's ``source coding theorem", it is always possible 
to encode the source in a way such that all sequences become  
equally probable ($H_{s}(\vec \sigma ) = const.$, non depending on the 
 $\sigma$'s). Source encoding reduces the redundancy 
in the source messages (not to be confused with ``channel 
encoding", see later). 

For reasons of simplicity, we will assume in the following that  the 
source has been encoded and that 
the noise is independent for any pair of bits (``memoryless channel"), i.e. 
$$ Q ( \vec u | \vec \sigma ) = \prod_{i} Q ( u_i | \sigma_i ) $$
In the case of a memoryless channel and a gaussian noise, Shannon 
calculated the channels capacity
$$ {\cal C } = { 1 \over 2 } \log_{2} (1 + { v^2 \over w^2 } ) $$
where $ v^2 / w^2 $ is the the signal to noise. In the weak signal to 
noise limit $ {\cal C } \sim v^{2} / (2 w^{2}  \ln2 )$ .

 Under the above assumptions, communication 
is a statistical inference problem. Given the transmission 
channel's output and the statistical 
properties of the source and of the channel, 
 one has to infer what message was sent. In order 
to reduce communication errors, 
one may introduce (deterministic) redundancy into the 
message (``channel encoding") and use this redundancy to 
infer the message sent through the channel (``decoding").
The algorithms which transform the source outputs to 
redundant messages are called error-correcting codes. 
 More precisely, instead of sending the $N$ original bits 
 $ \sigma_{i}  $, one sends $M$ bits $J_{k}^{in} $, $ k =1, \cdots , M $,
 $M > N $, 
constructed in the following way
$$ J_{k}^{in} \ = \ C^{(k)}_{i_{k_1}...i_{l_{k}}} \ \sigma_{i_{k_1}} \cdots 
\sigma_{i_{l_{k}}}    \eqno(1)  $$
where the ``connectivity" matrix 
$ C^{(k)}_{i_{k_1}...i_{l_{k}}} $ has elements zero or one. 
For any $k$, all the $ C^{(k)}_{i_{k_1}...i_{l_{k}}} $ except from one 
are equal to zero, i.e. the $J_{k}^{in} $ are equal to $ \pm 1$.
 $ C^{(k)}_{i_{k_1}...i_{l_{k}}} $ defines the code, 
i.e. it tells from which 
of the $\sigma$'s to construct the $k$th bit of the code. This kind of codes 
 is called parity checking codes because $J_{k}^{in} $ counts the 
parity of the minusis among the $ l_k $ $\sigma$'s. 
The ratio $ R = N / M $ which specifies the redudancy of the code, 
is called the rate of the code. 

We illustrate with a simple example of an $ R = 1/2 $ code. 
From the $ N $ $\sigma_i$'s we construct the $2N$ $J_{k}^{1,in} $, 
$J_{k}^{2,in} $, $i,k=1, \cdots , N $.
$$ C^{(1,k)}_{i_{k_{1}} i_{k_{2}} i_{k_{3}}} = \delta_{k,i_{k_{1}}+1}  
 \delta_{k,i_{k_{2}} }  \delta_{k,i_{k_{3}}-1} \ , \ \ 
 C^{(2,k)}_{i_{k_{1}} i_{k_{2}} i_{k_{3}}} = \delta_{k,i_{k_{1}}+1} 
 \delta_{k,i_{k_{3}}-1} $$
$$ J_{k}^{1,in} = \sigma_{k-1}  \sigma_{k}  \sigma_{k+1} \ , \ \ 
J_{k}^{2,in} = \sigma_{k-1}  \sigma_{k+1} $$ 

Knowing the source probability, the noise  probability, the code and 
the channel output, one has to infer the message that was sent.
The quality of inference depends on the choice of the code. 

According to the famous Shannon's channel encoding theorem, there exist 
codes such that, in the limit of infinitly long messages, 
it is possible to communicate error-free, provided the rate of the 
code $ R $ is less than the channel capacity $ {\cal C } $. This 
theorem says that such ``ideal" codes exist, but does not say 
how to construct them.

We will now show that there exists a close mathematical relationship 
between error-correcting codes and theoretical models of disosdered 
systems\ref{3,4,5,6}. 
As we previously said, the output of the channel is a sequence of $M$ 
 real numbers 
$ \vec J^{out} = \{ J_{k}^{out}, \ k=1, \cdots , M \} $, 
which are random variables, 
obeying the probability distribution $ Q(J_{k}^{out} | J_{k}^{in} ) $. 
 Once the channel output $ \vec J^{out}$ is known, 
it is possible to compute the probability $P( \vec \tau | \vec J^{out} ) $ 
for any particular sequence 
$ \vec \tau \ = \{ \tau_{i} , \ i=1, \cdots, N \} $ to be the {\it source } 
output (i.e. the information message).

More precisely, the equivalence between spin-glass models and 
error correcting codes is based on the following property\ref{5,6}.

{\it The probability $P( \vec \tau | \vec J^{out} ) $ for any sequence 
$ \vec \tau \ = \{ \tau_{i} , \ i=1, \cdots, N \} $ to be the information message, 
conditional on the channel output $ \vec J^{out} = \{ J_{k}^{out} , 
\ k=1, \cdots , M \} $ is given by  }
$$ \ln P( \vec \tau | \vec J^{out} ) \  = \ const \ - \  H_{s}(\vec \tau ) 
 \ +  \sum_{k=1}^{M} 
C^{(k)}_{i_{k_1}...i_{l_{k}}} \ B_{k} \ \tau_{i_{k_1}} \cdots 
\tau_{i_{l_{k}}}  \ \equiv \  - H_{t}(\vec \tau ) \eqno(2) $$
{\it where }
$$ B_{k} \ = \ B_{k} ( J_{k}^{out}) \ = \ {1 \over 2} \ \ln { Q(J_{k}^{out} | 1 ) \over 
Q(J_{k}^{out} | -1 ) }  \eqno(3) $$

We recognize in this expression the Hamiltonian of a p-spin spin-glass 
Hamiltonian. The distribution of the couplings is determined 
by the probability $Q(J^{out} | J^{in} )$. 

The proof is the following. The probability  
$ P( \vec \tau | \vec J^{out} ) $ for the source output to be  
$ \vec \tau $ when the channel output is $ \vec J^{out}$ is, by Bayes 
formula, 
$$ P( \vec \tau | \vec J^{out} ) \ = \ { P_{s} ( \vec \tau ) 
Q( \vec J^{out} | \vec J^{in} ) 
\over \sum_{ \vec \tau }   P_{s} ( \vec \tau ) 
Q( \vec J^{out} | \vec J^{in} ) } $$ 
where 
$J_{k}^{in} \ = \ C^{(k)}_{i_{k_1}...i_{l_{k}}} \ \tau_{i_{k_1}} \cdots 
\tau_{i_{l{k}}}    $.
Because the channel is memoryless and $J_{k}^{in} = \pm 1$, 
$$ \ln P( \vec \tau | \vec J^{out} ) \ = \ const. \ + \ 
\sum_{k=1}^{M} \ln Q(J_{k}^{out} | J_{k}^{in} ) \ + 
\ln P_{s} (\vec \tau )  $$ 
$$ \ln Q(J_{k}^{out} | J_{k}^{in} ) \ = \ { 1 \over 2}  
 \ln ( Q(J_{k}^{out} | 1 ) \  Q(J_{k}^{out} | -1 ) \ ) 
\ + \ { J_{k}^{in} \over 2} \ 
\ln { Q(J_{k}^{out} | 1 ) \over Q(J_{k}^{out} | -1 ) } $$
where $ const.$ means independent of $J^{in}$.
 To complete the proof, one has to substitute  the $J^{in}$'s 
according their definition as a product of the $ \tau $'s.

``Minimum error probability decoding" (or MED, see later), 
which is widely used 
in communications, consists in choosing the most probable 
sequence $ \vec \tau^0 $. This is equivalent 
to finding the ground state of the above spin-glass Hamiltonian.

In the case when 
$ Q(J^{out} | J^{in} ) \ = \ Q(-J^{out} | - J^{in} )   $
 (the case of a ``symmetric channel"), $ B_{k} ( J_{k}^{out}) = 
 - B_{k} ( - J_{k}^{out}) $ and one recovers the  invariance of the 
spin-glass Hamiltonian under gauge transformations
$ \tau_{i} \to \epsilon_{i} \tau_{i}, \quad 
B_{k} \ \to \  B_{k} \ \epsilon_{i_{k_1}} \cdots 
\epsilon_{i_{l_{k}}}, \quad \epsilon_{i} = \pm 1 $.

When all messages are equally probable and the transmission channel is 
 memoryless and symmetric, the error 
probability is the same for all input sequences. It is enough to compute 
it in the case where all input bits are equal to one. In this case, 
the error probability per bit $P_{e} $ is 
$ P_{e} \ = \ { 1 - m^{(d)} \over 2 }$, where 
$ m^{(d)} \ = \ { 1 \over N} \sum_{i=1}^{N} \tau_{i}^{(d)} $ and 
$ \tau_{i}^{(d)} $ is the symbol sequence produced by the decoding procedure. 

Let us  give a couple of examples of symmetric channels. 
 The first is the case of Gaussian noise 
(the ``Gaussian channel").
$$ Q(J^{out} | J^{in} ) = c \exp - {(J^{out} - J^{in} )^2 \over 2 w^2} ,
\quad B_{k} \ = \ {J_{k}^{out} \over w^2 }  \eqno(4) $$
The other example is when the output is again $ \pm 1$ (the 
``binary symmetric channel" ) 
 $$ Q(J^{out} | J^{in} ) \ = \ (1-p) \ \delta_{J^{out},J^{in} } \ + 
\ p \ \delta_{J^{out},-J^{in} } $$
$$ B_{k} \ = \ { \delta_{J_{k}^{out}, 1 } \over 2} \ \ln { 1 - p \over p } \  
\ + \ { \delta_{J_{k}^{out}, -1 }  \over 2} \ \ln {  p \over 1 - p } \ 
\ = \ { J_{k}^{out} \over 2} \ \ln { 1 - p \over p }  \eqno(5) $$
(the last equality holds because in this case $ J_{k}^{out} = \pm 1$).

Instead of considering the most probable instance, 
one may only be interested in 
the most probable value  $\tau_{i}^{p} $  of the ``bit" $\tau_i $\ref{7,8,9}.
 Because $\tau_i = \pm 1 $, the probability $p_i$ for $\tau_{i}^{p} =1 $ 
is simply related to $ m_i $, the average  of $\tau_{i}^{p} $,
$p_i = (1+m_i)/2 $. 

$$ m_i  \ = \ { 1 \over Z } \ \sum_{ \{ \tau_1 \cdots \tau_N \} } 
 \tau_{i} \ \exp -  H_{t} (\vec \tau )  \quad 
 Z = \sum_{ \{ \tau_1 \cdots \tau_N \} } 
  \ \exp -  H_{t} (\vec \tau )\quad \tau_{i}^{p}  = 
{\rm sign} \ ( m_i )  \eqno(6) $$
In the previous equation $m_i$ is obviously the thermal average 
at temperature $T=1$.
It is amusing to notice that for the gaussian channel or the binary
 symmetric channel, $T=1$ corresponds to Nishimori's temperature\ref{10}. 
Another amusing observation is that the so-called convolutional codes 
which are extremely popular in communications, 
correspond to one dimensional spin-glasses. Furthermore the decoding 
algorithm, which is called dynamical programming or ``Viterbi decoding 
algorithm", is nothing else than the transfer matrix algorithm of 
statistical mechanics. 

So the equivalence between parity checking error correcting codes 
and theoretical models of spin glasses is 
quite general and we have established the following 
dictionary of correspondence.
$$\eqalign { Error-correcting \ code \ &\iff \ Spin \ Hamiltonian \cr
Signal \ to \ noise \ &\iff \ J_{0}^2/\Delta J^2 \cr
Maximum \  likelihood \ Decoding\ &\iff \ Find \ a \ ground \ state \cr
Error \ probability \ per \ bit \ &\iff \ Ground \ state \ magnetization \cr
Sequence \  of \  most \  probable  \ symbols &\iff \ magnetization \  
 at  \  temperature \   T=1 \cr
Convolutional  \  Codes &\iff \ One \   dimentional  \  spin-glasses  \cr
Viterbi  \  decoding  &\iff \  Transfer \   matrix  \  algorithm } $$

This correspondence is not only an amusing mathematical curiosity, 
but can also be made useful by using the tools of modern statistical 
mechanics and the theory of disordered systems. 
Given a code, one can compute the error probability per bit if 
he is able to calculate the magnetization of the corresponding 
spin-glass model. There are at least two cases where this can be done. 

a) Week noise limit. Imagine first the case of no noise.
The minimal requirement for a good code is that the corresponding 
spin system has a unique ground state, well separated from the 
excited states by a finite energy gap. Consider next slowly 
switching on the noise. The energy levels become random variables 
whose probability distribution can eventually be computed. 
Error occurs when there is level crossing and a formerly excited 
state aquires a lower energy than the spin configuration which was 
the ground state in the absence of noise.
The probability of this to happen may be computed in certain cases. 

b)Extensive connectivity. This case corresponds to a mean field limit. 
To be precise we consider the case of a gaussian symmetric channel (i.e. 
gaussian noise) and a code defined by the following connectivity 
matrix $ C^{(k)}_{i_{k_1}...i_{l_{k}}} $ (see equation (1) );  
$ l_{k} = p $ for all $ k $ and $ C^{(k)}_{i_{k_1}...i_{p}} =1 $ 
for all possible $p$-spin multiplets. There are $ M = N!/(p!(N-p)!) $ 
 such multiplets. Therefore the rate of the code is 
$ R = N/M = p!(N-p)!/(N-1)! $. 
We consider the limit $ N \to \infty, p \to \infty, p^{2} / N \to 0 $ and 
$ p / \ln N \to \infty $. In this limit, the corresponding spin 
model is a slight generalization of Derrida's random energy model (REM)\ref{11}. 
This is easily seing, if one considers the case of all input bits 
equal to one. (This is not a loss of generality because all 
input sequences are obviously equivalent when the noise is 
symmetrically distributed around zero.) Derrida considered the 
case of gaussian random couplings with zero average and standard deviation 
$ \Delta J^{2} =W^2 $. The only difference with the present case 
is that the coupling average is $J_0 =V $ where $ V^2 $ is the signal 
power which is non zero.
 (In fact it can be shown that a Gaussian noise is not required. Only the 
 first two moments of the noise distribution are relevant, i.e. 
the computation is valid not only for Gaussian noise but also for more 
general symmetric noise distributions.) 

For the spin model to have a nontrivial thermodynamic limit, we 
consider the case of a signal power such that $ V= v p! / N^{p-1} $ and 
a noise power 
$ W^2 =w^2 p! / N^{p-1} $, $p$ and $N \to \infty $, while $v$ and $w$ 
are kept fixed. 
 The signal to noise power ratio is then 
 $ V^2 /  W^{2}  =  v^2 p! / w^{2} N^{p-1}  $
Using arguments \`a la Derrida or ``replica" calculations, 
(neither of these arguments is rigorous) 
it can be shown that, in this model,  
  the ground state magnetization is 
$m =1 $ for $ v^{2} /w^2 > 2 \ln 2  $ and zero otherwise. 
As we saw above, $m =1 $ means zero error probability per bit 
for the corresponding code.  The above inequality $ v^{2} /w^2 > 2 \ln 2  $ 
is equivalent to $R < C$. In other words, the error-correcting code, 
corresponding to the the random energy model, is an ideal code, 
i.e. allows error-free communication if $R < C$. 
One may wonder how fast this code approaches the asymptotic regime. 
It turns out that it is possible to compute the asymptotic expansion 
of $m$ as $p \to \infty $. This is done by using the ``replica method".
 The result of this computation is
that for $v^{2} /w^2 \sim 2 \ln 2 $,
$$ m=1-{\exp(-p v^{2} /w^2) \over \displaystyle \sqrt {p} }\ 
\  c ( v^{2} /w^2 ) \qquad  \qquad
c( 2 \ln2 )=.987  $$

To the best of my knowledge, the only other explicitly known ideal 
codes are pulse position modulation (or ppm) codes. They can briefly 
be described as follows. During a time interval $T$, one can transmit 
one of $ N $ possible symbols. $T$ is divided into $ N $ 
subintervals of duration $ \delta = T/N $. To send the $i$'th symbol, 
one sends during the $i$'th time subinterval an electric pulse of 
duration $ \delta $ and amplitude $h$. $ \delta $ and $ h $ have to be 
chosen depending on the noise power and the desired reliability. 
It can be shown that in the limit $ h \to \infty $ and 
$ \delta \to 0 $, this code is ideal. 
Both the 
REM and ppm codes become ideal in the limit of infinite redundancy 
and zero signal to noise power. 

Up to now we only considered parity checking codes, for which the ``alphabet" 
has length 2, i.e. there are only two symbols, $ \sigma = 1 $ and 
$ \sigma = - 1 $ 
Let me finally mention that many of the previous results can 
be generalized\ref{5,6} to the case of an 
alphabet of length $l$. One may establish a one to one correspondence 
between the $l$ symbols of the alphabet and the elements of 
a finite group with the same number of elements. Spin multiplication is 
replaced by group multiplication. These codes can be seeing as 
an interpolation between parity checking codes and pulse position 
modulation codes.

\vfill\eject

\vglue .7 true cm
\centerline{\bf References }
\vglue .6 true cm

\item {1)} Shannon, C. E., A Mathematical Theory of Communication, 
{\it Bell Syst. Tech. J.} 27, 379 and 623 (1948)

\item {2)} Shannon, C. E. and Weaver W., A Mathematical Theory of Communication,
 (Univ. of Illinois Press, 1963)

\item {3)} Sourlas, N., {\it Nature} 339, 693 (1989)

\item {4)} Sourlas, N., in {\it Statistical Mechanics of Neural Networks}, 
Lecture Notes in Physics 368, ed. L. Garrido, Springer Verlag (1990)

\item {5)} Sourlas, N., Ecole Normale Sup\'erieure preprint (April 1993)

\item {6)} Sourlas, N., in {\it From Statistical Physics to 
 Statistical Inference and Back,}
 ed. P. Grassberger and J.-P. Nadal, Kluwer Academic (1994) p. 195.

\item {7)} Ruj\'an, P., {\it Phys. Rev. Lett.} 70, 2968 (1993)

\item {8)} Nishimori, H., {\it J. Phys. Soc. Jpn. }, 62, 2973 (1993)

\item {9)} Sourlas, N., {\it Europhys. Lett.} 25, 169 (1994)

\item {10)} Nishimori, H., {\it Progr. Theor. Phys.} 66, 1169 (1981)

\item {11)} Derrida, B., Random-energy model: An exactly solvable model 
of disordered systems, {\it Phys. Rev.} B24, 2613-2626  (1981)

\end